# HomeGuard: A Smart System to Deal with the Emergency Response of Domestic Violence Victims

**Anik Islam[1], Arifa Akter[2] and Bayzid Ashik Hossain[3]**

[1] **Department of Science & Information Technology, American International University of Bangladesh Banani, Dhaka, Bangladesh**

[2] **Department of Science & Information Technology, American International University of Bangladesh Banani, Dhaka, Bangladesh**

[3] **Department of Science & Information Technology, American International University of Bangladesh Banani, Dhaka, Bangladesh**

**Abstract**

Domestic violence is a silent crisis in the developing and underdeveloped countries, though developed countries also remain drowned in the curse of it. In developed countries, victims can easily report and ask help on the contrary in developing and underdeveloped countries victims hardly report the crimes and when it's noticed by the authority it's become too late to save or support the victim. If this kind of problems can be identified at the very beginning of the event and proper actions can be taken, it'll not only help the victim but also reduce the domestic violence crimes. This paper proposed a smart system which can extract victim's situation and provide help according to it. Among of the developing and underdeveloped countries Bangladesh has been chosen though the rate of reporting of domestic violence is low, the extreme report collected by authorities is too high. Case studies collected by different NGO's relating to domestic violence have been studied and applied to extract possible condition for the victims.

## 1. Introduction

Every day we see different kinds of violence like murder, rape, hijack etc. Some of them get their proper punishment or some of them don't. But most of them at least get a chance to file a report against it. There is another kind of violence named domestic violence that remains silent in our society and is increasing day by day. Domestic violence happens when a member of family, spouse, husband or ex-partner tries to physically or psychologically dominate or harm the victim. There are many forms of domestic violence or abuse including physical abuse or assault, sexual abuse, financial and career control, use of religious doctrine, emotional abuse, stalking, threats, blackmailing etc. [1]. Domestic abuse can happen to anyone of any folk, gender, religion, age or sexual orientation. It can also happen to the couples who are married or dating or living together. It can happen to the employee who engaged in house chores. But statistics describe that 95% victims of domestic violence or abuse are female and 90% of the delinquents are male [18]. The major intention of domestic violence is to gain control or maintain dominance over the victim. There are multiple forms for abuse like Physical, Psychological, and Sexual etc. Physical abuse is an intentional act to cause feelings of injury, pain or other physical sufferings by way of physical or bodily contact. It includes hair pulling, beating with objects, shaking, pushing, choking, kicked, poisoned, slapped, punching, burning or any other actions that cause any form of physical injury to the victim. Psychological abuse also is known as emotional abuse or psychological violence or mental abuse. It's another form of violence that occur from another person and cause mental trauma including anxiety, depression or stress. Though it doesn't leave scars on bodies, it can cause a great negative effect on confidence and self-respect of the victims. On the contrary, the intention for abuse related to sexual is quite different. The main culprit of sexual abuse are the people with distorted mentality. Sexual abuse is a sexual activity without victims' consent using force or making threats or blackmailing or giving false promises. Most of the victims and delinquents know each other. Even they can be close relatives of each other. Any sexual activity with children performed by the adult or older child against or with the consent of victims also will treat as sexual abuse known as child sexual abuse. This a silent violence that starting to unfurl its wings silently among the society. The effect of abuse can't be described in a word. Sometimes effect of abuse can be little or sometimes its effect draws the victim to the death. Effects can be divided into two categories such as psychological and Physical. Major victims of psychological abuse commonly reports high amount of stress, anxiety and fear. Depression is also very common in the domain of psychological abuse as perpetrators made the victim feel that he/she is the reason behind the abuse and he/she provoked him/her to do that and the victim also gets acute criticism from the perpetrators. Statistics showed that, 60% of victims meet the criteria for depression after or during the termination of the relationship which increases the risk greatly of suicidal tendencies [3]. There are some mentionable short term effects of Psychological abuse like shame, guilt, frequent crying, aggression, avoidance of eye contact etc. The most commonly reported psychological effect of domestic violence is Post-Traumatic Stress Disorder (PTSD). Post-Traumatic Stress Disorder (PTSD) is related to mental health which triggered by an appalling event, either witnessing or experiencing it. The long-term effects of physical abuses are chronic pain, pelvic pain, ulcers, migraines and arthritis [19]. Pregnant victims experience the risk of miscarriage, injury or death of fetus and preterm labor [2]. Victims may develop ways to cope with the trauma that left from physical abuse or perhaps a victim forgets the nightmares but the trauma left the effects like eating disorder, self-harm, drug addiction, trouble in sleeping, discomfort of touching even if it's the touch of a parent. The effect of physical abuse basically related to the minor or major physical injuries like fractures, burns, bruises, head injuries etc. which requires medical attention and hospitalization [2]. Domestic violence is no less from other kinds of violence. Statistically, Women become the victim of different form of abuses more than men. Though it may seem that domestic violence only occurs in the developing and underdeveloped countries but statistics shows that no country is free from domestic violence. Findings from an analysis of 10 separate domestic violence appearance studied by the Council of Europe showed that 1 in 4 women experience domestic violence during their lifetime and between 6−10% of women face domestic violence in a year [4]. Between one and five million women in the USA become the victim of abuse at the hands of their intimate partner [5].

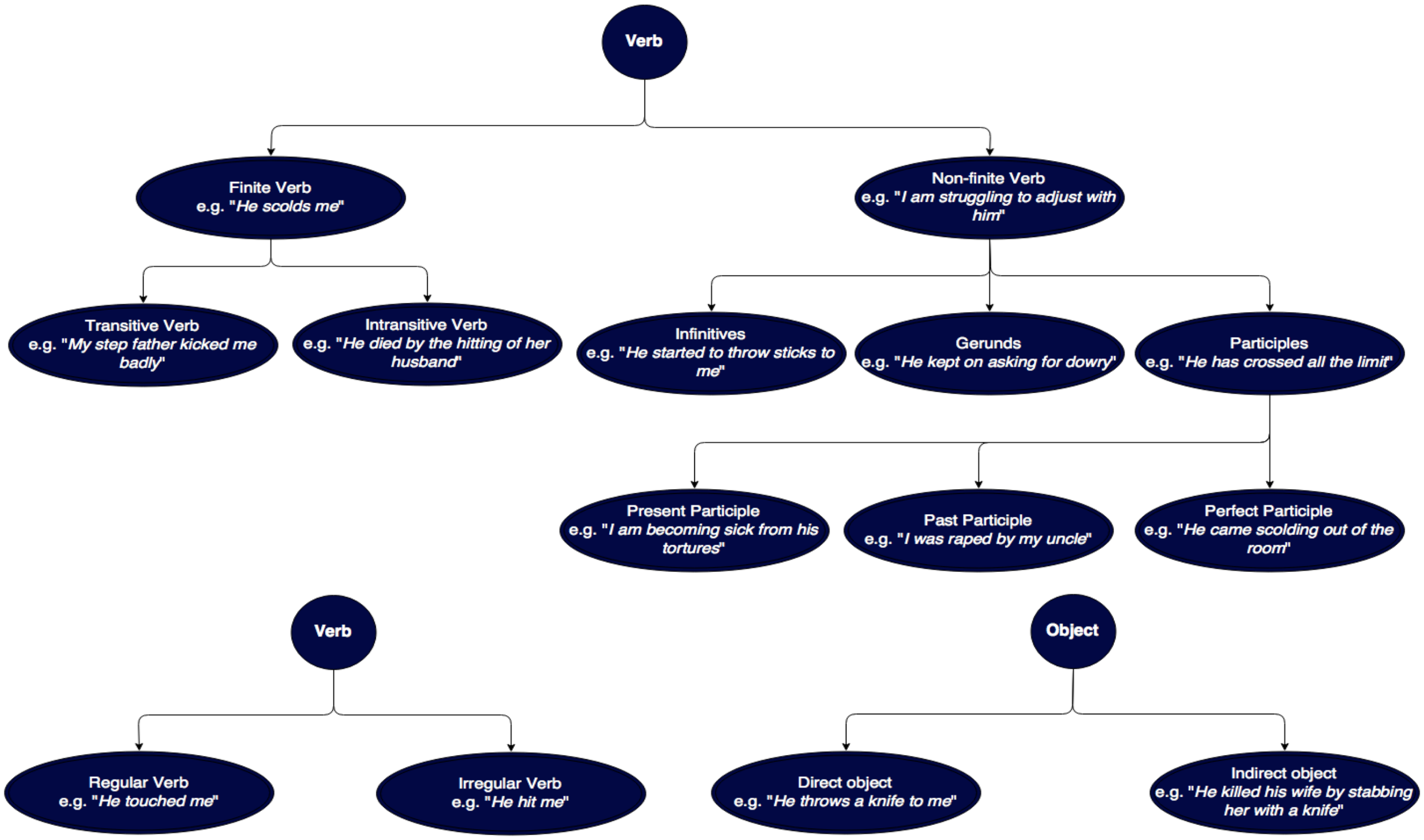

Fig. 1 Classification of verb with examples.

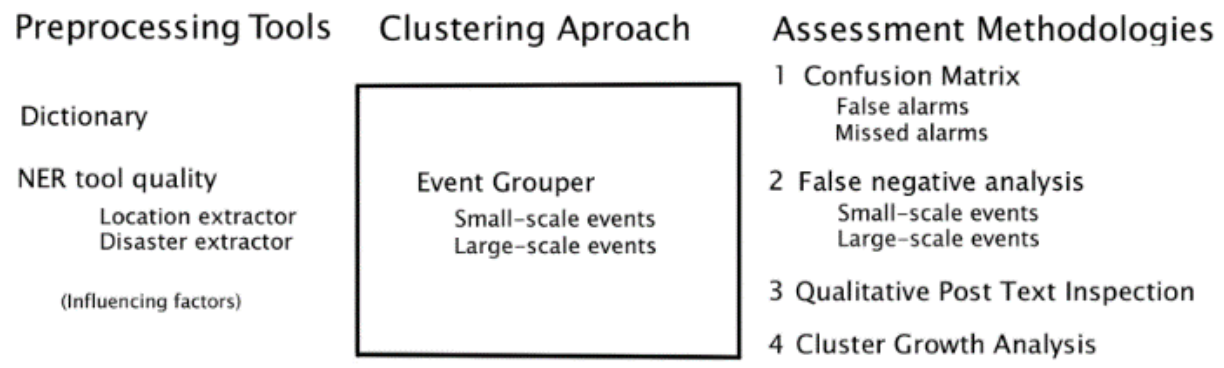

Fig. 2 Research model proposed in Klein et al.

In developed countries, majority of the abuses are reported directly or indirectly by the victim but in developing countries and underdeveloped countries, majority of the victims tolerate domestic abuses silently. Victims remain silent during facing abuses are related to social, fear etc. like fear of humiliation in the society, fear of the losing child, lack of family support etc. Whatever the motive is, the ultimate outcome is suicide or death caused by the intimate partner. As this is not reported at early stage, the perpetrator get the courage to continue abusing by increasing the limit. If no steps being taken soon or aid being provided, it'll increase dramatically day by day. To reduce domestic violence and provide aid to the victims, we propose a smart system named "HomeGuard". The main goal of this project is to build an intelligent system using Natural Language Processing (NLP) Techniques and Ontology in which victims can send their messages by stating the conditions of their situation and system will read these messages and detect the crime level of the perpetrator and provide aid and support to the victim according to it as soon as possible.

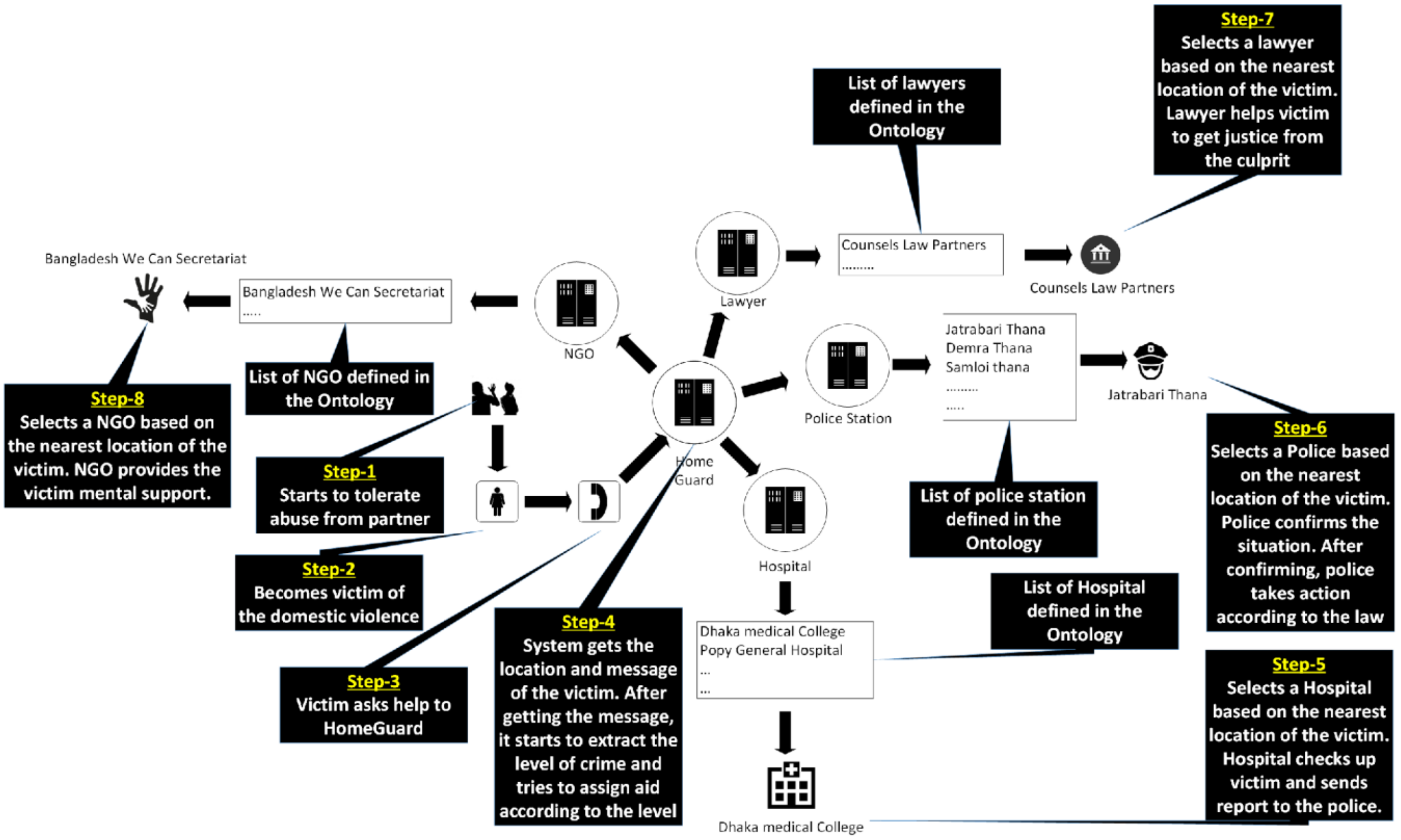

Fig. 3 System structure of HomeGuard.

The remaining sections of this paper is organized as follows. Section 2 presents the related work. Section 3 provides the text processing mechanism of the proposed system along with the description about the ontology and overall process of the system. Section 4 describes about the experiment and development. Finally, Section 5 concludes the article with future works.

## 2. Related Work

In this section, we discuss existing researches regarding the system which deals with the emergency situation. Several researchers had worked on emergency system based on data extracted from tweets but no one proposed any system that provide help to the victims according to our knowledge. However, there are some similarities between our HomeGuard and other researches in which they proposed emergency based system in the twitter stream. The main goal of that researches and our proposed system is to extract emergency situation from the data and provide help to the victims. Among of the proposed emergency based models [8] and [7] are highly notable. Both proposals are based on emergency detection from tweets but with different approaches.

In [8] they proposed an approach to detect emergency events. To annotate tweets, they performed discourse analysis on tweets from their selected four data sets. Their data sets contains

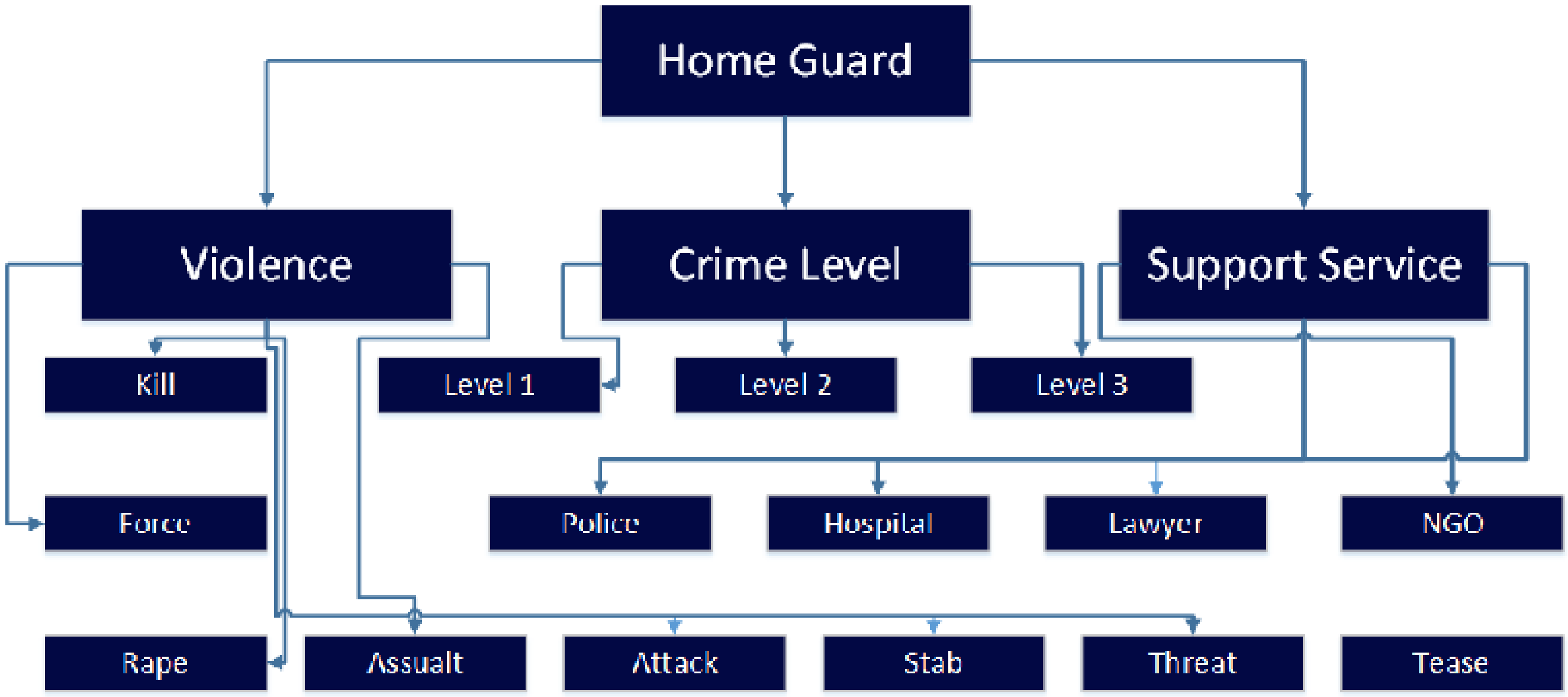

Fig. 4 Class structure of HomeGuard

• **Situational Awareness Tweet Content** (Providing information related to situation).

• **Objective Tweet Content** (Providing purely factual information).

• **Subjective Tweet Content** (Providing tweets contain sentiment or opinion).

• **Personal/Impersonal Tweet Content** (Providing tweets regarding the situation attached by the twitterer).

They proposed a model which extracts feature from the tweets. Majority of the tweets contains symbols and artifacts which can confuse the classification process. So they replaced symbols and artifacts with a unique symbol. After that, they have tokenized the tweets into words and calculate each words frequency. They have used POS tagger to tag tweets.

The list of features they have added to the SA Classifier is provided below:

1. Unigrams and their raw frequency.
2. Bigrams and their raw frequency.
3. Part-of-speech tags.
4. Subjectivity of tweets (Objective/Subjective).
   (a) Hand-annotated
   (b) Predicted by the subjectivity classifier.
   (c) Derived by the OpinionFinder ([10], [13]).
5. Register of tweet (Formal/Informal)
   (a) Hand-annotated
   (b) Predicted by register classifier.
6. Tone of tweet (Personal/Impersonal)
   (a) Hand-annotated
   (b) Predicted by classifier.

In [7], they considered different type of natural and human disasters which includes weather-related disasters like hurricanes, fires, flooding and also the geological disasters like earthquakes or health related epidemics. The complexity of disasters varies with respect to the scale, dynamics, and spatial distribution. Small-scale disasters cover a very small area and effect on people. On the contrary, large-scale disasters cover a very large area and the huge number of people can become affected by it and this make it easy to detect the situation as people would report about the situation. Tweets are collected via external API and after collecting tweets, they filtered the twitter stream manually with the selected keywords like ***911***, ***Aid***, ***Ambulance***, etc. They have limited the geographic scope of detecting emergencies by putting observation ranges through the Twitter API. 10% of the retrieved messages represent the complete communication and this makes the number collected tweets very large in order to cover the actual event as much as possible. It's very important to separate emergency messages from non-emergency messages via effective filtering process to avoid false alarm. A surveys [8] showed that degree of extracted information from some tweets are strongly correlated with the sentiment which they filtered in their model. They defined the classification as follows

**Definition of Emergency Classification. Given t an emergency taxonomy, they defined a message m belonging to the domain of emergency if more then n words exists where** $\forall x = \{1...n\} w_x \in t \wedge w_x \in m$.

They defined an emergency identifier as follows:

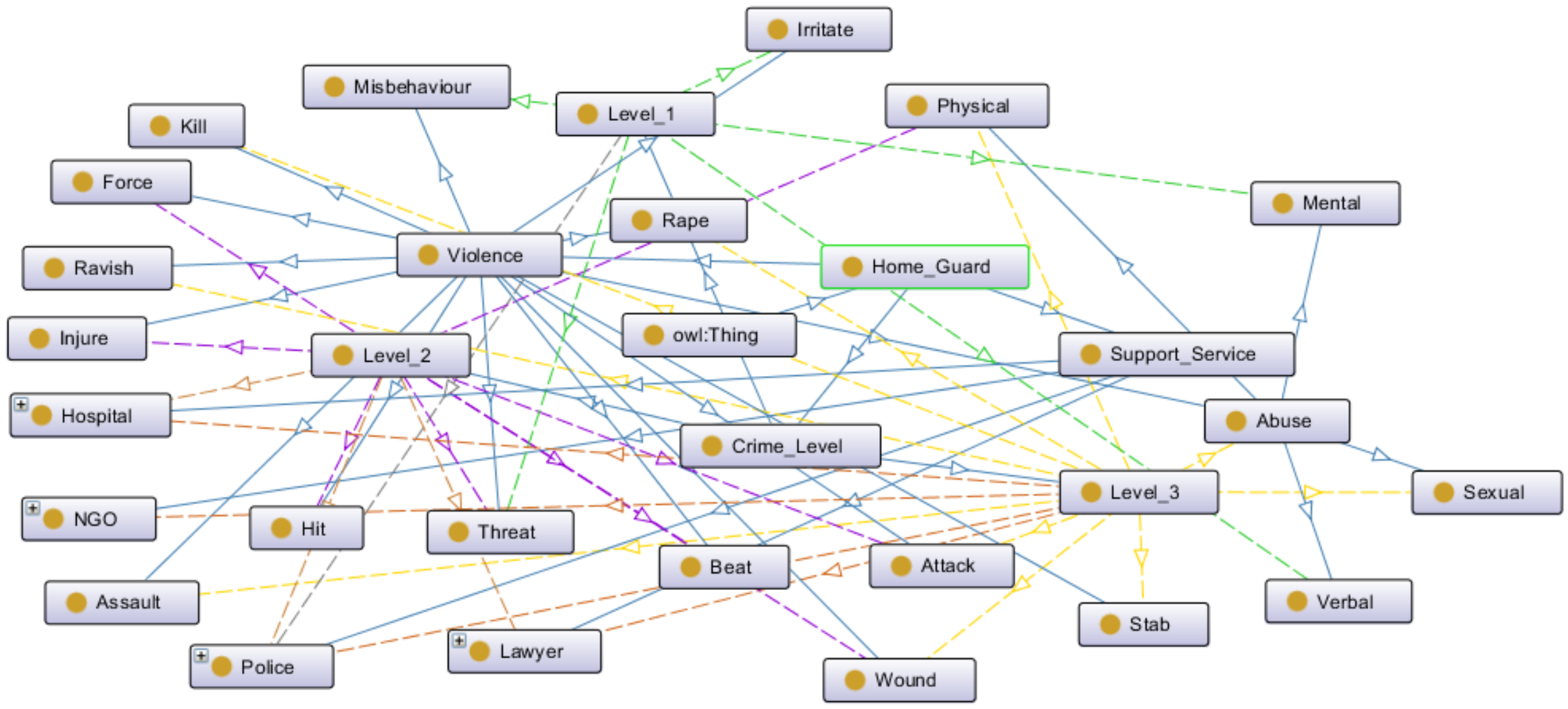

Fig. 5 Ontology structure of HomeGuard

***Definition of Emergency Identification:* An emergency identifier is built after following syntax:**

$$\langle\textbf{emergency identification}\rangle :=$$

$$\left\{\begin{array}{l}\langle\textbf{emergency keyword}\rangle + \langle\textbf{disaster name}\rangle \vee \\ \langle\textbf{emergency keyword}\rangle + \langle\textbf{location hierarchy}\rangle\end{array}\right\}$$

$$\textbf{where} \quad \langle\textbf{location hierarchy}\rangle := \left\{\begin{array}{l}\langle\textbf{country}\rangle + \\ \langle\textbf{region}\rangle + \\ \langle\textbf{city}\rangle + \\ \langle\textbf{district}\rangle\end{array}\right\}$$

Their model is showed in Fig 2. In their model, they automatically removed tweets with punctuation or several letter or other misspellings. Finally, summery of their approaches, they collected chunks through the twitter crawlers using different keywords which includes both large and small scale of events. They have developed a tagging tool which helps to classify emergency post. After the classification, each message is assigned under a predefined specific emergency event.

## 3. Our Proposed System: HomeGuard

In this section, we introduce our proposed model named HomeGuard. In our proposed system, victim sends emergency message to the system. Then the system reads these messages and provides help to the victim according to their emergency level. Message from a victim can contain one or more sentences. A sample message is provided below

***My husband is an unemployed and has no will to work. That's why I have to work outside of the home. Every night he hit me for money. Last month he broke my hand by beating with a rod. Now I need help badly.***

The goal of the system is to extract level of crime that occurred to the victim. In best cases, first sentence may contain the information regarding the level of crime and in worst case the last sentence may contain the information regarding the level of crime. That's why every sentence is equally important in the process. Sentences from victims can be three types which mentioned below:

- **He hit me**
- **He didn't hit me**
- **He will hit me**

The first sentence displays actual action faced by the victims but the second sentence is a negative sentence which has no effect on the victim and the third sentence displays uncertainty which victim feels that he/she may get beat by the perpetrator or it's just victims imagination. The system won't act on last two types of sentences as these sentences lacks the actual event occur by the perpetrator. Some sentences are questions like this

***What should I do to stop him?***

The system will also ignores question type sentences as this type of sentences lacks the action of the perpetrator or situation of the victims. If we come to the tense of the situation, three tenses can be considered which are

- **Present tense** {My husband punches me on my face}
- **Past tense** {My husband punched me on my face}
- **Future tense** {My husband will punch me one day}

Present tense and past tense proves that the action has happened and victim needs help but future tense brings uncertainty to the fact. So system won't process future tenses. However there can be two types of sentences in the message and they are

- **Simple sentence**
- **Compound sentence**

Simple sentences are sentence with one clause. On the contrary, compound sentences are sentences with multiple clauses connected by connectives. Compound sentences can be like this

- $\langle\textbf{Main clause}\rangle\langle\textbf{Connective}\rangle\langle\textbf{Main clause}\rangle$ {He first scolded me and after that, he started to beat me}
- $\langle\textbf{Main clause}\rangle\langle\textbf{Connective}\rangle\langle\textbf{Subordinate clause}\rangle$ {He started to hit me when I asked for money}
- $\langle\textbf{Subordinate clause}\rangle\langle\textbf{Connective}\rangle\langle\textbf{Main clause}\rangle$ {After the marriage, he started to beat me for dowry}

Main clauses contain a subject and verb and make sense on their own on the other hand Subordinate clause also contains a subject and verb but it depends on the main clause to express any meaning and connectives which joins the clauses can be adverbs, conjunctions and prepositions. The basic structure of messages which system will get like this

$$\langle\textbf{Home Guard}\rangle := \langle\textbf{Victim's Message}\rangle$$

$$\langle\textbf{Victim's Message}\rangle := \langle\textbf{Sentences}\rangle$$

$$\langle\textbf{Sentence}\rangle := \langle\textbf{Simple Sentence}\rangle \vee \langle\textbf{Compound Sentence}\rangle$$

$$\langle\textbf{Compound Sentence}\rangle :=$$

$$\{\langle\textbf{Main Clause}\rangle \wedge \langle\textbf{Connective}\rangle \wedge \langle\textbf{Main Clause}\rangle\}$$

$$\vee\{\langle\textbf{Main Clause}\rangle \wedge \langle\textbf{Connective}\rangle \wedge \langle\textbf{Subordinate Clause}\rangle\}$$

$$\vee\{\langle\textbf{Subordinate Clause}\rangle \wedge \langle\textbf{Connective}\rangle \wedge \langle\textbf{Main Clause}\rangle\}$$

$$\langle\textbf{Connective}\rangle := \langle\textbf{Conjunction}\rangle \vee \langle\textbf{Preposition}\rangle \vee \langle\textbf{Adverb}\rangle$$

$$\langle\textbf{Simple Sentence}\rangle := \langle\textbf{Clause}\rangle$$

The system will focus more on verbs than other parts of speech in the sentence as only verbs represent the action. Fig 1 shows many form of verbs those exists in the grammar. As we have described different kinds of verb, each verb has different types of behavior. To extract information, we have used NLP tools. NLP is one of the popular components of Artificial Intelligence (AI) and is widely used to understand human language. Part-Of-Speech tagging (POS Tagging), Chunking, Named Entity Recognition (NER) and Semantic Role Labeling (SRL) are the most used among the NLP techniques [12]. In HomeGuad, extraction of actual action from each behavior of verb is important as aid will be generated from the level of action. Among the techniques of NLP, POS Tagger tags parts of speech to the messages but it doesn't consider the context of the sentence. After getting word tagged by POS tagger, we processed verb with different kinds of behavior. Verb may not be in the present form after retrieving from the corpus. So we lemmatized the word to get the actual form of verb. The extracted word may be a synonyms of a word defined in the data model. By using site WordNet, we mapped synonym problems. This will help to reduce redundancy and defining new properties in the data model. "Ontology" term comes from the field of philosophy. In the field of philosophy, it means the theory of being but in the field of artificial intelligence, its meaning is different [22]. Ontologies are specifications of conceptualizations which is often thoughts as a directed graph in which concept as a node connected by edges which represent the relation between nodes [14]. The notion of concepts which defined in knowledge representation understood as a set of objects or individuals [15]. As a basis for the sharing of knowledge, ontology has been widely used in Artificial Intelligence and Information Science [21]. In ontologies concepts often presented their nodes with corresponding natural language concept names. The relations in the backbone structure of the ontology are "is-a" whereas other remaining structure may include relations like "located-in", "part-of", "is-parent-of" etc. [16]. OIL (Ontology Inference Layer), DAML (Darpa Agent Markup Language) + OIL and OWL (Web Ontology Language) are used in developing ontologies [20]. Ontology is heterogeneous and light weight [11]. Moreover any rules or constrains can easily be applied on the classes and objects. It is also supported in multi platforms. Ontology is also easily modifiable without hurting the data. The Ontology structure is provided in the Fig 5. In the Fig 4, we have mentioned our class structure. We have designed our class like at top most class is the Home Guard Class, after that there are three sub classes of it which are Violence, Crime Level, and Support Services. Each subclass has its own sub classes like Violence has multiple sub classes like kill, threat, attack etc. Crime level class has level 1, level 2 and level 3 sub classes and Support service class has police, hospital, lawyer sub classes. Police, Hospital, Lawyer and NGO has data properties like name, address, phone number, latitude and longitude. Sub classes of Crime level class are connected with the sub classes of violence class via object properties as crime level will be defined according to violence. Sub classes of Support service class are connected with the crime level as support service class will be delivered according to the level of crime is occurring. The overall structure and working process of HomeGuard is provided in Fig 3. Finally, the processes that follows to extract situation and provide help to the victim via Home Guard includes:

Table 1: Victim's message to the system and the services that will be provided to the victim

| |
|---|
| **Message:** My husband come home drunk and hit me every day. I need help.<br><br>**Service found:** Hospital, Lawyer, Police |
| **Message:** My uncle raped me. Nobody here in my home. Please help me.<br><br>**Service found:** Hospital, Lawyer, Police, NGO |
| **Message:** My step mom doesn't love me. She gets angry with me for little things. She doesn't give me food properly and hit me every day. I want to get rid of this situation. Please help me.<br><br>**Service found:** Hospital, Lawyer, Police |
| **Message:** She used to regularly scream at me and hit me, but when I needed stitches in my head after she had attacked me with a knife while drunk.<br><br>**Service found:** Hospital, Lawyer, Police |
| **Message:** After Nabila had threatened me with a knife on more than one occasion, and I'd successfully ducked missiles, she finally got her aim right one morning and hit me with a bowl about one centimeter from my eye. I turned up to work that morning with blood-stained clothing and had to explain my fragile situation.<br><br>**Service found:** Hospital, Lawyer, Police |
| **Message:** He stabbed me in my hand multiple time by a knife during our fight. Though he felt sorry about this and asked my forgiveness, I don't think that he won't repeat this. Please help!<br><br>**Service found:** Hospital, Lawyer, Police, NGO |
| **Message:** I was threatening to leave my boyfriend because he told me he went to the strip club- and got a private dance from one of the girls. I was angry I was screaming, "I'm going to leave you" my bf snapped and said "YOU FUCKING WAIT HERE IM COMING BACK" he went to the garage, grabbed thick rope and black tape and used these to tie up my hands, feet and mouth. I tried to fight him away from me but he kept attacking me and I couldn't move. I couldn't breathe properly because I was so scared. He left me in the room and I was alone- I kept crying I was struggling to tell myself "it will be over soon don't worry" after two hours he come back and placed a bowl of water on the ground near me to taunt me and tease me. I could hear his stupid friend laughing which upset me more. Five hours.. When he finally came to his senses and untied me- my wrists and feet were marked and bleeding. He's like "are you going to leave me now? I did this to teach u a lesson you stupid bitch" I said "I want to go home" he's like "you better not leave me otherwise it's going to get worse" I said "how" he's like "test me and you'll see". This continues for every fight. Please help me!<br><br>**Service found:** Hospital, Lawyer, Police |
| **Message:** We had an argument, and she suddenly escalated and started to hit me repeatedly.<br><br>**Service found:** Hospital, Lawyer, Police |

Table 1: Victim's message to the system and the services that will be provided to the victim

| |
|---|
| **Message:** He came upstairs and asked me to get out of bed to help him look for a work shirt. I didn't get out of bed. I replied that I wanted to go to sleep. He suddenly turned on me. He hit me, somehow got me in the position of being flat on my back. He stood on me and spat in my face.<br><br>**Service found:** Hospital, Lawyer, Police |
| **Message:** He hit me in my stomach and kept knocking my head. I kept trying to push him away but he wouldn't stop. This is how I had my first miscarriage.<br><br>**Service found:** Hospital, Lawyer, Police |
| **Message:** He always punched me on the tops of my legs or in the head. he only hit me there because no one would see the bruises that way. Help!<br><br>**Service found:** Hospital, Lawyer, Police |
| **Message:** My boyfriend gave the threat to kill me. He always told me it was my fault. HELP!<br><br>**Service found:** Police |
| **Message:** The violence became a regular occurrence in our relationship as a husband-wife. One time, he broke my finger and I had to see a hand specialist. On another occasion, I went to A & E with bruising to my head, face and body. He spat at me, pushed me, kicked me and bit me. Once he even tried to run me over. I can't bear this anymore.<br><br>**Service found:** Hospital, Lawyer, Police, NGO |
| **Message:** I am married for 4 years. Every now and again she'd flip out during a fight and slap or slug me in the jaw, I only ever fought back once to stop her from hitting me (punched her in the arm), which immediately brought everything to a standstill and the "you just hit me" argument started up.<br><br>**Service found:** Police |
| **Message:** I am young, so is my wife. She began hitting me when things didn't go her way, which became pretty often; I wanted to leave but we had a kid together.<br><br>**Service found:** Hospital, Lawyer, Police |
| **Message:** My ex roommate had some serious issues with his girlfriend. They'd split up at one point and she used a spare key she had (made a copy before giving his key back to him) to break into his house and walk into his bedroom while he was sleeping. She stood over him and then took her lanyard with keys attached and began beating him. He woke up being hit over and over with keys being swung around on a nylon rope. She was screaming at him while doing it. He threw her out of the house and she threatened to call the police to report that he'd started the physical portion of the fight and saying she hit him back in self-defense. What he should do?<br><br>**Service found:** Hospital, Lawyer, Police |

1. Split message into sentences.
2. Filter out special character from the sentence like @, #, ! etc.
3. Filter out the question type sentence.
4. Filter out sentence which contain uncertain actions.
5. Filter out the negatives sentences (How do I explain it to him?).
6. Filter out the question type sentence.
7. Tag sentence with pos tagger.
8. Extract action by analyzing verb which tagged from pos tagger with its behavior.
9. Lemalized the extracted action to get actual action name.
10. Find out the maximum action level that sentence contain.
11. Extract crime level based on the action from ontology.
12. Extract help service information based on the crime level from ontology.
13. Assign services to the victim.

## 4. Development and Evaluation

The development of HomeGuard consists of two parts. First part is the text processing and another is the data modeling. HomeGuard used c# version of the Stanford CoreNLP library[1] to process victim's messages. It used Log-linear Part-Of-Speech Tagger of Stanford CoreNLP library [17] and processed actions according to our dataset of different kinds of verb. In the data modeling part, our proposed system used Protege[2] with RDF for ontology and used dotNetRDF[3] for processing RDF on the server, which is a library for processing RDF in c#. To make query in Ontology, system used SPARQL (SPARQL Protocol and RDF Query Language)[4] in the dotNetRDF library. The service related data is collected and tested on the basis of Bangladesh. The efficiency depends on the training sets of action processing system. The more training set is added to the system, the more efficient the processing will be. The result of sample messages of victims is provided in the Table 1. The result is shown based on the data defined in the Ontology. According to the Table 1, the first message of a victim contains a violence term named "hit" which is the level 2 crime in the eye of the Ontology. For level 2, hospital, lawyer, and police are tagged with the crime. So, victim will get the help of these services as soon as possible. For the second message in the Table 1, the message contains a violence term "raped" which is the level 3 crime in the eye of the Ontology. For level 3, hospital, lawyer, police, and NGO are tagged with the crime. Like this, other messages will be processed in the similar way. The target of this system is to extract highest level of violence that a victim faced and provide help according to it. If the message contains multiple levels of crime, the system will always pick the maximum level.

## 5. Conclusions

We have designed, developed and evaluated HomeGuard, a smart intelligent system which will provide help to victims in critical situation. The situation is detected from the statement of victims which they send to the system. To improve system intelligence, it requires more training sets containing the behavior of actions. As we have implemented on the circumstance of Bangladesh including its help services and definition of the crime, system is unknown to the other country's situation and crime patterns but the system can adjust any type of dataset providing from the victim with proper training set. As a further improvement, we plan to gather other developed and underdeveloped countries data and their crime patterns to make the system universal. We also plan to implement the system in the local language instead of English so that every level of people can ask help in the system and get help from it.

### Acknowledgments

This research was partially supported by the Department of Science & Information Technology, American International University of Bangladesh (AIUB). The authors are grateful for this support.

[1] http://stanfordnlp.github.io/CoreNLP/
[2] http://protege.stanford.edu/
[3] http://dotnetrdf.org/
[4] https://www.w3.org/TR/rdf-sparql-query/

**Anik Islam** received his Bachelor degree in Software Engineering in 2014 from American International University of Bangladesh (AIUB). He first joined Bengal Solutions Ltd. as a software engineer in 2012. Later, he joined Dynamic Software Ltd. in 2014. After that, he joined Next IT Ltd. as a software engineer. He took leave from Next IT Ltd. in order to peruse his Master degree and Ph.D. Currently, he is doing Masters in Computer science at American International University-Bangladesh (AIUB). Side by M.Sc., he is working in the Bangladesh branch of Frenclub Mobile Ltd. as a part-time team leader in the Mobile Application development team. He is seeking for Ph.D. opportunity in Computer Science related fields. His research interest includes intelligent system, semantic web, data science, human computer interaction (HCI), distributed system, internet of things (IoT), web of things (WoT) and big data.

**Akter Arifa** received her Bachelor degree in Computer Science and Software Engineering in 2014 from American International University-Bangladesh (AIUB). In 2014, she joined Rupshi Sweaters Ltd. as IT Manager. Currently, she is doing Masters in Computer science at American International University of Bangladesh (AIUB). Her research interest includes intelligent system, semantic web, distributed system, data mining and big data.

**Bayzid Ashik Hossain** is currently working as an assistant professor at Department of Computer Science in American International University-Bangladesh (AIUB). Previously he was working at Esolution Europe S.r.l as Software Engineer in Milan. He has completed Masters in Computer Science from University of Trento, Italy. His specialization was Software Technologies. He completed his undergraduate in Computer Science and Information Technology from Islamic University of Technology (IUT), Dhaka, Bangladesh. His research interest includes knowledge representation and management, semantic web, intelligent systems and data science.